\documentclass[sigconf]{acmart}
\usepackage{blindtext}
\usepackage[dvipsnames]{xcolor}

\AtBeginDocument{%
  }

\setcopyright{cc} %

\setcctype{by}

\acmConference[MuC'26]{Mensch und Computer 2026 – Tagungsband, Gesellschaft für Informatik e.V.}{30. August - 02. September 2026}{Duisburg, Germany}
\acmDOI{10.18420/muc2026-mci-ws102-257}

\begin{document}

\title[It is not enough to give your moderation rules to ChatGPT]{It is not enough to give your moderation rules to ChatGPT}
\subtitle{Policy-as-Prompt Moderation and Its Potential Impacts on Community Governance} %

\author{Anna Neumann}
\email{neumann@rc-trust.ai}
\affiliation{
    \institution{Research Center for Trustworthy AI}
    \institution{University of Duisburg-Essen}
    \country{Germany}
}
\orcid{0009-0000-9672-8087}

\author{Jasmin Wyss}
\orcid{0009-0009-9144-9105}
\affiliation{%
  \institution{Ruhr-University Bochum}
  \country{Germany}
}

\author{Ivy Turk}
\orcid{0000-0002-4705-4749}
\affiliation{%
  \institution{Ruhr-University Bochum}
  \country{Germany}
}
\orcid{0009-0009-9144-9105}

\author{Rebekah Overdorf}
\orcid{0000-0003-3462-9539}
\affiliation{%
  \institution{Research Center for Trustworthy AI}
  \institution{Ruhr-University Bochum}
  \country{Germany}
}

\renewcommand{\shortauthors}{Neumann et al.}

\begin{abstract}
Content moderation practices and governance paradigms are changing rapidly, as fewer human moderators are deployed as `experts' by social media companies in a \textit{centralized} manner. Instead, the companies are focusing more on community approaches, relying on volunteers to provide accurate information and make correct decisions. In \textit{decentralized} moderation, communities have always relied on volunteers, updated community guidelines, and internal discussions thereof. For both content moderation paradigms, Artificial Intelligence (AI) seems like it could help ease moderation burdens of time, mental health, and accuracy.
One possible way to operationalize AI in content moderation is a ``policy-as-prompt'' approach, where the policy is formulated as a natural-language prompt and then passed to a large language model (LLM). This model then aids in moderation tasks. In this paper, we briefly lay out the technical and governance properties of this approach, and argue that its limitations lead to specific risks and harms that have to be addressed. Towards alleviating them, we lay out multiple considerations towards more effective prompt governance, but ultimately find that writing prompts alone is not appropriate for ensuring meaningful community governance.
\end{abstract}

\keywords{Artificial Intelligence (AI), System Prompts, Content Moderation, Prompt Governance, Policy-as-Prompt, Generative AI}

\maketitle

\section{The (New) Content Moderation Landscape}

Artificial intelligence has had a large impact on the content moderation landscape. Machine learning classifiers are now the first line of moderation on almost all major social media platforms, such as TikTok, Facebook, Instagram, or X/Twitter, while human moderation has been scaled back or removed from many of them. In modern moderation, humans mostly act as reviewers of edge cases, appeals reviewers, and (increasingly) training data labellers.~\cite{hunsberger_llm_nodate, kuo_unsung_2023, newton_secret_nodate, noauthor_scroll_nodate}
This is in part due to a shift of trust and safety policies being politically contested~\cite{alizadeh_content_2022}, as X/Twitter and Facebook/Instagram (Meta), for example, have reduced moderation enforcement in the name of free speech~\cite{bbcMetaReplace, verfassungsblogMuskTechbrocracy}. 

Large Language Models and Large Multimodal Models (LLMs \& MLLMs) are increasingly proposed and trialed for tasks related to content moderation, promising to alleviate pressures on time and mental health while outperforming human moderators~\cite{palla_policy-as-prompt_2025, yew_dynamic_2026}. 
For example, Meta's \textit{Llama Guard}~\cite{inan2023llamaguard} takes a safety taxonomy as an input and then labels posts as `safe' or `unsafe'. More recently, OpenAI published an open-weight model \textit{gpt-oss-safeguard}~\cite{openai25gptosssafeguard}, where developers supply a written policy and the model reasons over it to produce a labelled and explained decision on a post. 

Due to scaling (and monetary hurdles), biases, and performance drops in non-English text, these models are not yet widely independently deployed~\cite{franco_integrating_2025, gomez_algorithmic_2024, shahid_think_2025, yew_dynamic_2026}. As models get better, these problems are expected to become negligible. LLMs would then swiftly replace multiple tasks in content moderation workflows. Still, one underlying problem would persist: simply giving an LLM a policy or a set of rules as a prompt alone cannot govern its behavior absolutely~\cite{neumann2026prompt}, and will have implications on content moderation practice and governance. Therefore, in this paper, we explore the implications of setting moderation policies through natural language prompts given to AI systems on community governance.

\section{Content Moderation as a Governance Problem}

Content moderation has never been just about the application of a set of textual rules: it is also a community governance practice~\cite{gorwa20algorithmic, alizadeh_content_2022, klonick_new_2017}. Moderators deliberate and discuss edge cases; formalized appeals processes determine how community members can seek recourse. Additionally, mechanisms to ensure meaningful transparency of implemented rules are vital~\cite{norval_disclosure_2022, suzor_what_2019}.
The process of moderation also needs to be able to adapt to the evolution of norms in the moderated space, leaving room for changing definitions, contexts, and understandings to emerge~\cite{franco_integrating_2025}. These cannot be described by one centralized way of knowing, as the same words (can) carry different meaning across communities and contexts: ``Words change depending on who speaks them; there is no cure''.~\cite{nelson2015argonauts} 

These sense-making operations have historically been facilitated through human processes, but with the advent of generative AI systems paradigms in moderation, practices like deliberation by multiple moderators over specific edge cases could be outsourced. Therefore, depending on how LLM-based moderation is implemented, it risks disempowering communities by destructing their influence and recourse options on moderation practices. %

\section{Moderation through Policy-by-Prompt}

The seemingly most accessible approach to integrating LLMs into moderation is called ``policy-as-prompt''. The moderation policy itself is encoded as a natural-language instruction
~\cite{palla_policy-as-prompt_2025}. This prompt is directly supplied as a system instruction~\cite{wallace_instruction_2024} to a general-purpose LLM, which is asked to apply it to a given piece of content. Policy-as-prompt is appealing, as moderation could be reconfigured by simply editing a sentence in the prompt instead of retraining a model. Thus, it promises rapid adaptation to emerging harms, new policies by organizations or jurisdictions, or a shift in platform or community norms. Yet, the swiftness of adaptation is dictated by interdisciplinary writing processes of the new prompts as well as their evaluations and trials before deployment. However, as outlined in previous research~\cite{neumann2026prompt}, system instructions are not reliable enough to give governance guarantees towards goals such as alignment, performance, or robustness of the system. 

An additional limiting factor of these approaches is the presence of what are sometimes described as ``prompt stacks''~\cite{neumann_who_2026, wallace_instruction_2024}. Prompts are processed according to a hierarchical order that prioritizes those set by foundation-model developers before downstream developers, API users, or interface end-users. This prioritization means that moderation instructions (e.g., asking the LLM to apply rules to a specific piece of content) ought to only be processed if they align with the input from prompt layers before. Higher priority prompts should not be able to be overridden by lower priority ones~\cite{neumann_position_2025}. Therefore, what looks like writing hard rules is, in practice, adding one instruction to a possibly unaligned hierarchy.%

For generative AI approaches in content moderation, the mixture of technological limitations~\cite{salini_sarcasm_2023, kumar_watch_2023}, well-known evasion or circumvention tactics (e.g., prompt injections)~\cite{perez_ignore_2022, greshake_not_2023}, insufficient guardrail effectiveness~\cite{dong_safeguarding_2025, bertollo_breaking_2025}, and other behavior stemming from scaffolding around the foundation model~\cite{neumann_ai_2026, zhan_injecagent_2024}, ultimately means that policy-as-prompt approaches on their own are not stable enough to deliver on the envisioned guarantees. 
As prompt governance unfolds further, prompts have to be supplemented with robust evaluations and other governance measures to have a chance at easing content moderation burdens. Still, the seeming ease of implementation and oversight incentivizes the application of policy-as-prompt.

\section{Possible Impacts of Policy-as-Prompt}

In addition to prompt-based moderation having specific technical risk and harm possibilities, the removal of human expertise and accountability from moderation processes will be of consequence. We trace both kinds of impacts across two moderation paradigms. 

\subsection{Centralized Moderation}
On platforms where moderation is centralized, e.g., done by a governing body, expertise plays a vital role. Policy specialists draft guidelines that then get tested and adjusted with the help of moderators and domain experts~\cite{palla_policy-as-prompt_2025, ruckenstein_re-humanizing_2020}, ideally in dialogue with community members~\cite{schaffner_community_2024}. This process shapes understanding of the policy that moderators then have to apply in practice. With policy-as-prompt regimes, a deliberated policy is compressed into a single artifact translated not for a human moderator but for an AI system. As \citet{palla_policy-as-prompt_2025} put it: ``Writing for machines relies on machine interpretation''. New AI policy translation is therefore not focused on translating between organizations and communities, but between organizations and AI. The AI prompt now carries all of the nuance once passed through levels of judgment and competing incentives~\cite{west_raging_2017}. This translation additionally might not work as intended due to limitations of language operationalization of stochastic systems in general~\cite{neumann2026prompt, palla_policy-as-prompt_2025}. %

\subsection{Decentralized Moderation}

Self-governing communities rely heavily on volunteers and a contextual understanding of their community. Volunteer moderators write their own guidelines, deliberate over them in (mostly) transparent ways, and local norms evolve with the community and their discussions~\cite{cook_commercial_2021}. A prompted moderator AI 
could change community dynamics: the rules could still be written, revised, and understood by the community, then defined in a (possibly non-transparent) prompt set and applied by AI. Some moderators may not want to disclose instructions, because of worries about `easier circumvention'~\cite{neumann_who_2026}.
This also changes rule enforcement dynamics. When a moderator from the community applies a rule, this is based on context and knowledge of the community. By no longer performing this task, the community members lose vital expertise. %
This would impact the feedback loop in self-governance and potentially lead to %
less effective and situated moderation. The rules may even be changed in their formulation to account for misinterpretations by AI~\cite{yakura_empirical_2025, adkins_what_2024}. %
The community guidelines become what the generative AI system can operationalize and change in parallel to the development of LLMs~\cite{neumann2026prompt, kim_natural_2026}. Change in community norms will then (in part) be governed by the affordances of technological developments.

\section{Considerations on Policy-as-Prompt Practices}

Policy-as-prompt approaches already exist, and are likely to increase across organizations due to perceived ease of intervention and accessibility~\cite{neumann2026prompt}. As such, some limitations and implications should be taken into consideration when designing such systems. 

First, given the technical limitations and idiosyncrasies of generative AI models, with the possibility of prompt injection and (emergent) circumvention methods, they should generally be handled with caution in content moderation~\cite{zeng_how_2025, kumar_watch_2023, hartmann_lost_2025}. 
As such, if policy-as-prompt is implemented, it needs to go beyond specifying words to go into system instructions. Writing a prompt and giving it to a general-purpose model %
is simply not robust enough at steering system behavior~\cite{neumann2026prompt}. Prompts need to be evaluated on their impact on model behavior and judgment, as well as their effects across the prompt stack, e.g., through sensitivity analyses. 

Second, technical evaluation of policy-as-prompt will still not give governance guarantees. The approach needs to be embedded in robust governance structures that can deliver meaningful disclosure, agency, and accountability~\cite{norval_disclosure_2022, neumann_who_2026, lewicki_out_2023} to the governed communities.

Third, the inability of generative AI systems to take responsibility for their outputs limits the content moderation tasks they can be deployed for. Moderation actions should be contestable and/or appealable. As it stands, LLMs cannot take the required accountability for their (moderation) actions. Therefore, they should only \textit{assist} in human decisions, or \textit{supplement} human deliberation processes.

\section{Conclusion}

Although AI might have its place in content moderation, using policy-as-prompt on its own %
is not enough. Moderation is highly dependent on the fluidity of language, and, thus, on context, time, and intention. As such, content moderation necessitates various sense-making operations, which include deliberation, appeals, and contextual interpretation of broader community values. Outsourcing these mechanisms (even partially) to LLMs has a profound impact. Not only because they take some of these sense-making operations out of human hands, but also because LLM ``decisions'' are based on a warped mirror of past language and its context, instead of %
current use of language within the specified context. As such, looking forward requires acknowledging that AI can only look backward, which both justifies and necessitates human-driven processes in content moderation.

\bibliographystyle{ACM-Reference-Format}
\bibliography{bibliography}

\end{document}